\documentclass[letterpaper,english,reprint, prb, reprint, aps]{revtex4-2}
\usepackage[T1]{fontenc}
\usepackage[latin9]{inputenc}
\usepackage{verbatim}
\usepackage{amsmath}
\usepackage{amssymb}
\usepackage{graphicx}
\usepackage{float}
\usepackage{bbold}
\usepackage{xcolor}

\makeatletter

\pdfpageheight\paperheight
\pdfpagewidth\paperwidth

\usepackage{braket}

\newsavebox{\savedimage}

\makeatother

\usepackage{babel}
\begin{document}
\title{Fate of multipolar physics in $5d^2$ double perovskites}
\author{Ahmed Rayyan$^{1}$, Xiaoyu Liu$^{2}$, Hae-Young Kee$^{1,3}$}
\email{hy.kee@utoronto.ca}

\affiliation{{\footnotesize{}$^{1}$Department of Physics, University of Toronto,
Toronto, Ontario M5S 1A7, Canada}}
\affiliation{{\footnotesize{}$^{2}$Department of Materials Science and Engineering, University of Washington, Seattle, Washington 98195, USA}}
\affiliation{{\footnotesize{}$^{3}$CIFAR Program in Quantum Materials, Canadian
Institute for Advanced Research, Toronto, Ontario M5G 1M1, Canada}}
\date{\today}
\begin{abstract}
In a cubic environment, the ground state of spin-orbit coupled $5d^2$ ions is a non-Kramers $E_g$ doublet, which hosts quadrupole and octupole moments. A series of $5d^2$ osmium double perovskites Ba$_2M$OsO$_6$ (M = Mg, Ca, Zn, Cd) have recently been proposed to exhibit multipolar orders.  
We investigate the structural properties of these materials using $\textit{ab}$-$\textit{initio}$ calculations and find that the cubic structure is unstable for the Cd compound while the Mg, Ca, and Zn materials retain $Fm\bar{3}m$ symmetry. We show that Ba$_2$CdOsO$_6$ favours a rhombohedral $R\bar{3}$ structure characterized by $a^-a^-a^-$ octahedral tiltings as indicated by unstable $\mathcal{T}_{1g}$ phonon modes. 
Trigonal distortions split the excited $T_{2g}$ triplet into an $E'_g$ doublet and an $A_g$ singlet, which may cross energy levels with the $E_g$ doublet and suppress the multipolar physics. 
We find a window where $E_g$ remains the lowest energy state under trigonal distortion, enabling the emergence of multipole phases in non-cubic crystal environments.

\end{abstract}
\maketitle

\section{Introduction}

Strongly correlated materials feature many distinct phases often classified by different order parameters associated with their broken symmetries. 
A textbook example is the magnetic order described by various arrangements of magnetic dipole moments.
The magnetic order parameter can be probed using several experimental techniques such as neutron scattering, and its onset would accompany thermodynamic phase transitions. 
Under certain conditions, the dipole moment is absent yet higher-rank moments may transition into an ordered state.
However, due to their multipolar nature, they can be ``hidden'' from some experimental probes \citep{tripathi2007}.
A physical framework encompassing these cases motivates studying the ordering mechanisms of states with non-trivial multipolar moments.

The $f$-electron systems have been a natural platform to explore multipolar physics, as the rare-earth ions carry total angular momentum $J$ via
strong spin-orbit coupling (SOC) \citep{santini2009,Kuramoto2009}. 
Examples include the $4f^{2}$ Pr materials where the Pr$^{3+}$ ions carry a $J=4$ multiplet \citep{sakai2011,ito2011,onimaru2011,sato2012,tsujimoto2014,konic2021}. 
The ninefold degeneracy is lifted by octahedral or tetrahedral crystal electric fields (CEFs) and yields a doublet where the Kramers degeneracy does not apply due to the even number of electrons. 
The resulting non-Kramers doublet lacks a dipole moment yet carries quadrupole and octupole moments. 
%
%
A Landau theory for Pr$^{3+}$ 1-2-20 materials has been developed in recent years \citep{hattori2014,hattori2016,lee2018,ishitobi2021} and it suggests that the hidden octupole moment can be revealed within magnetoelastic experiments by applying a $[111]$ magnetic field \citep{patri2019}.

The situation in $d$-electron systems is more subtle. 
In the $3d$ Mott insulators, the light magnetic ions carry a much weaker SOC so that the orbital angular momentum is often quenched by CEFs. 
In the case where the orbital degrees of freedom may yet fluctuate, such as for one hole or one electron in $e_g$ states (e.g. $3d^9$ or low-spin 3$d^7$ respectively), then orbital ordering is usually found via the Kugel-Khomskii mechanism \citep{Kugel1982}.
The resulting orbital ordering is equivalent to a motif of charge quadrupoles and can be accompanied by a structural transition via the cooperative Jahn-Teller effect \citep{gehring1975}. The ordering of higher multipoles, including octupolar moments, is difficult to achieve in the lighter transition metals. 

However, higher-rank multipoles may be relevant in the heavier transition metal compounds such as the $5d^1$ and $5d^2$ double perovskites (DPs) $A_{2}BB'$O$_{6}$ \cite{Chen_2011}. 
The combination of strong SOC, large separation between magnetic $B'$O$_{6}$ octahedra, and high cubic symmetry satisfies the necessary preconditions for local multipolar physics \citep{chen2010,Chen_2011,dodds2011,ishizuka2014,vasala2015,romhanyi2017,svoboda2021}.
A promising platform for octupolar ordering lies within the $5d^{2}$ DPs, where the magnetic $B'$ ion features a $J=2$ SOC multiplet that is split into a low-lying non-Kramers doublet $\left(E_{g}\right)$ and an excited triplet $\left(T_{2g}\right)$ via orbital $t_{2g}$-$e_{g}$ mixing \citep{Voleti_2020,doi:10.7566/JPSJ.90.062001}. 
Similar to the $4f^{2}$ case, the non-Kramers $E_{g}$ doublet carries multipolar moments with interactions obtained by projecting $d$ orbital hopping channels onto the $E_{g}$ doublet to form an effective pseudospin-1/2 model. 
These include compass quadrupole and Ising octupole interactions \cite{Khaliullin_2021} where the strength (and sign) of these terms depends on the details of each material under consideration \cite{Churchill_2022}.

The $5d^2$ barium osmate DPs with Ba$_{2}$$M$OsO$_{6}$ and $M\in\left\{ \text{Mg, Ca, Zn, Cd}\right\}$ have been proposed as a series of compounds which features multipolar orderings \citep{YAMAMURA2006,Thompson_2014,Marjerrison_2016,Maharaj_2020}.
These materials exhibit at most a single transition at temperature $T^{*}$ that does not coincide with a reduction of cubic symmetry. 
Various candidates for the ordering observed in these materials include antiferro-quadrupolar \citep{Khaliullin_2021}, ferri-octupolar \citep{lovesey2020}, and ferro-octupolar \citep{Voleti_2021,Pourovskii2021,Mosca2022,Churchill_2022}.
The octupolar order hypothesis is favoured for the Mg, Ca, and Zn compounds, as the ordering at $T^{*}\sim30-50$ K coincides with the onset of muon-spin relaxation ($\mu$SR) oscillations, signalling a loss of time-reversal symmetry \citep{Thompson_2014,Marjerrison_2016}. 
On the other hand, the Cd compound does not feature thermodynamic anomalies or $\mu$SR oscillations down to $T^{*}=0.47$ K \citep{Marjerrison_2016}, and its low-temperature structural properties is yet to be determined by the high-resolution synchrotron x-ray diffraction as in Ref. \citep{Maharaj_2020}.
Thus, a set of questions arise naturally which we aim to resolve in this work: is the cubic structure of the $5d^2$ Os DPs Ba$_{2}$$M$OsO$_{6}$ stable at low temperatures, and if not, what is the fate of the multipolar physics in the low-symmetry environment?

The paper is organized as follows. 
In Section \ref{sec:reviewJ2} we briefly review how the $E_g$ doublet arises out of the $J=2$ states and discuss challenges in identifying the crystal structure in some 5$d$ DPs. 
In Section \ref{sec:StructuralOs} we investigate the phonon spectrum of each compound using \emph{ab}-\emph{initio} density functional theory (DFT) simulations, and find that the Cd compound favours a rhombohedral structure characterized by a set of octahedral tiltings.
In Section \ref{sec:trigDistort} we evaluate how the $E_g$ and $T_{2g}$ states are modified in the rhombohedral structure where each OsO$_{6}$ octahedron is trigonally compressed along the $[111]$ direction. 
We conclude with a summary of our findings and their implications, as well as avenues of future work. %

\section{$E_g$ and $T_{2g}$ States in $5d^{2}$ Osmium Double Perovskites\label{sec:reviewJ2}}

\begin{figure}
\includegraphics[scale=0.17]{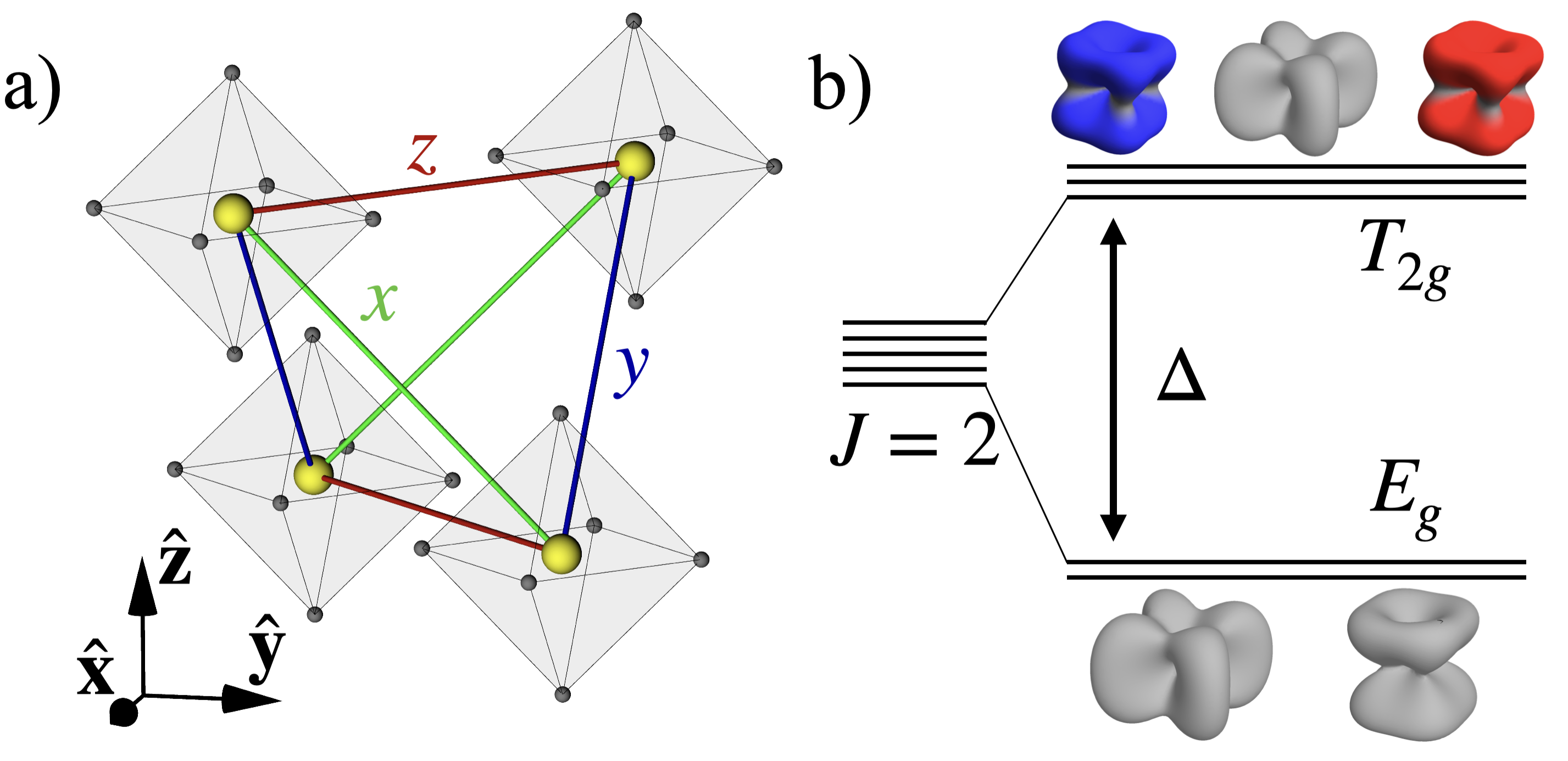}\caption{\label{singleIonCoord} a) Os$^{6+}$ ions (yellow) enclosed in oxygen octahedral cages (grey) arranged in the fcc structure. 
The crystallographic $xyz$ coordinate system is shown along with fcc nearest-neighbour bonds of type $x$ (green), $y$ (blue), and $z$ (red). 
The $M$O$_{6}$ octahedra and BaO$_{12}$ cuboctahedra are not shown. 
b) The single-ion level scheme for $5d^2$ electrons in an octahedral CEF, where the low-lying $E_{g}$ doublet is separated from the excited $T_{2g}$ triplet by a residual CEF splitting $\Delta\sim10$ meV.
The corresponding $E_{g}$ and $T_{2g}$ states are shown, with regions of red and blue denoting non-zero spin density. {[}Reproduced from Fig. 1b) in Ref. \citep{Rayyan2023}.{]}}
\end{figure}

The $5d^2$ Os DPs Ba$_{2}$$M$OsO$_{6}$ form a rock-salt pattern of alternating $M$O$_6$ and OsO$_6$ corner-sharing octahedra that enclose Ba$^{2+}$ ions. The Os$^{6+}$ ions carry the relevant magnetic degrees of freedom and form an fcc lattice, see Fig. \ref{singleIonCoord}a).
The octahedral CEF lifts the electronic $d$ orbital degeneracy into irreps of the octahedral group $O_h$ as $\Gamma_{l=2}=t_{2g}\oplus e_g$, with the low-lying $t_{2g}$ triplet separated from the excited $e_{g}$ doublet by energy gap $10Dq$. 
The two $t_{2g}$ electrons form a high-spin configuration with $S=L=1$ via Hund's rules. 

The strong SOC $-|\lambda|\mathbf{L}\cdot\mathbf{S}$ of the $5d$ ions with $|\lambda| < 10 Dq$ favours a total $J=2$ multiplet with magnetic moment $\mathbf{M}=-\mathbf{L}+2\mathbf{S}=\mathbf{J}/2$ with magnitude $\sim1.25\mu_{B}$ \citep{Chen_2011}. However, heat capacity, $\mu$SR, and inelastic neutron scattering experiments find a low-lying doublet separated from excitations by a gap of $\Delta\sim10$ meV with a small magnetic moment of $\mathcal{O}(0.1\mu_{B})$, in apparent contradiction with the $J=2$ picture \citep{Marjerrison_2016,Maharaj_2020}.
Some mechanisms for a residual CEF which generates a low-lying $E_{g}$ doublet and an excited $T_{2g}$ triplet as shown in Fig. \ref{singleIonCoord}b) have been proposed to lift the $J=2$ degeneracy, including $t_{2g}$-$e_{g}$ mixing via SOC and Hund's coupling, or non-spherical Coulomb interactions in a $t_{2g}$-only model \citep{Voleti_2020}. 
Defining $\ket{m}\equiv\ket{J=2;J^{z}=m}$ where $m\in\left\{ 2,1,0,-1,-2\right\} $, the $E_{g}$ and $T_{2g}$ states in the cubic harmonic basis are linear combinations of the states shown in Fig. \ref{singleIonCoord}b), and are given by
\begin{align}
\ket{\uparrow} & =\frac{1}{\sqrt{2}}\left(\ket{-2}+\ket{2}\right), & \ket{T_{x}} & =\frac{i}{\sqrt{2}}\left(\ket{-1}+\ket{1}\right),\nonumber \\
\ket{\downarrow} & =\ket{0}, & \ket{T_{y}} & =\frac{1}{\sqrt{2}}\left(\ket{-1}-\ket{1}\right),\label{eq:EgT2gcubic}\\
 &  & \ket{T_{z}} & =\frac{i}{\sqrt{2}}\left(\ket{-2}-\ket{2}\right).\nonumber 
\end{align}
The $E_{g}$ doublet is of non-Kramers type and has vanishing dipole moments $\braket{\mathcal{P}_{E_{g}}^{\dagger}\mathbf{J}\mathcal{P}_{E_{g}}}=0$ where $\mathcal{P}_{E_{g}}=\sum_{\omega\in\left\{ \uparrow,\downarrow\right\} }\ket{\omega}\bra{\omega}$ projects onto the $E_{g}$ states. 
Three higher-rank multipoles retain a finite moment: the quadrupole operators $Q_{z^{2}}=\frac{1}{\sqrt{3}}\left(3J_{z}^{2}-\mathbf{J}^{2}\right)$ and $Q_{x^{2}-y^{2}}=J_{x}^{2}-J_{y}^{2}$, and the octupole operator $T_{xyz}=\frac{\sqrt{15}}{6}\overline{J_{x}J_{y}J_{z}}$ where the overline symbol denotes symmetrization \citep{santini2009}. These multipolar operators have the same matrix elements as the Pauli matrices within the $E_g$ doublet and can be considered as effective pseudospin-1/2 operators. Microscopic models of the $E_g$ doublets feature a variety of multipolar-ordered ground states including antiferro-quadrupolar (AF$\mathcal{Q}$) and ferro-octupolar (F$\mathcal{O}$) orders \citep{lovesey2020,Khaliullin_2021,Voleti_2021,Pourovskii2021,Mosca2022,Churchill_2022}. The lack of an observed structural transition for the Mg, Ca, and Zn compounds seems to imply that AF$\mathcal{Q}$ is unlikely to be the low-temperature phase. On the other hand, the predicted antiferro-distortions tends to generate tiny structural deformations which can be missed in some diffraction experiments.

For example, consider the $5d^{1}$ DP Ba$_{2}$MgReO$_{6}$ \citep{sleight1962compounds,BRAMNIK2003235} which features a $j_{\text{eff}}=3/2$ pseudospin on each Re$^{6+}$ ion and two thermodynamic anomalies at temperatures $T_{q}>T_{m}$ where $T_{q}=33\text{ K}$ $\left(T_{m}=18\text{ K}\right)$ corresponds to the onset of quadrupolar (dipolar) order \citep{marjerrison2016d1,hirai2019,hirai2020,arima2021}. In Ref. \citep{marjerrison2016d1} the $\mu$SR data suggests that there are two inequivalent oxygen sites which hints at an underlying tetragonal distortion, yet no such effect is detected via neutron diffraction. On the other hand, a very small cubic-to-tetragonal distortion below $T_{q}$ was detected using the high-resolution synchrotron x-ray diffraction on a sample with high crystallinity, and is predominantly associated with the ordering of alternating $Q_{x^2-y^2}$ quadrupoles near $T_q$ \citep{hirai2020}. 
This highlights the challenges faced in the accurate structural determination of DPs, especially the detection of quadrupolar-induced antiferro-distortions in $5d$ materials \citep{vasala2015,hirai2020}. It is also worthwhile to note that in the case of CeB$_6$ ($4f^1$) there is no change in B$_6$ positions despite the established AF quadrupole order \citep{nakao2001, tanaka2004direct, fazekas99}. 

While the lack of a structural transition in the Ba$_2M$OsO$_6$ for the $M=$ Mg, Ca, Zn compounds has been verified using synchrotron x-ray diffraction \citep{Maharaj_2020}, it is worthwhile to examine the stability of the high-symmetry cubic structure. Note that the non-Kramers doublet degeneracy is sensitive to general structural deformation which may result in the loss of multipolar physics.
Furthermore, the situation regarding the $M=$ Cd material is not yet fully determined. In the next section we investigate the phonon spectra 
of the four $5d^2$ Os DPs to examine the stability of the cubic structure.
%
\section{Structural Properties of $5d^{2}$ Osmium Double Perovskites\label{sec:StructuralOs}}
\begin{figure}
\centering{}\includegraphics[scale=0.26]{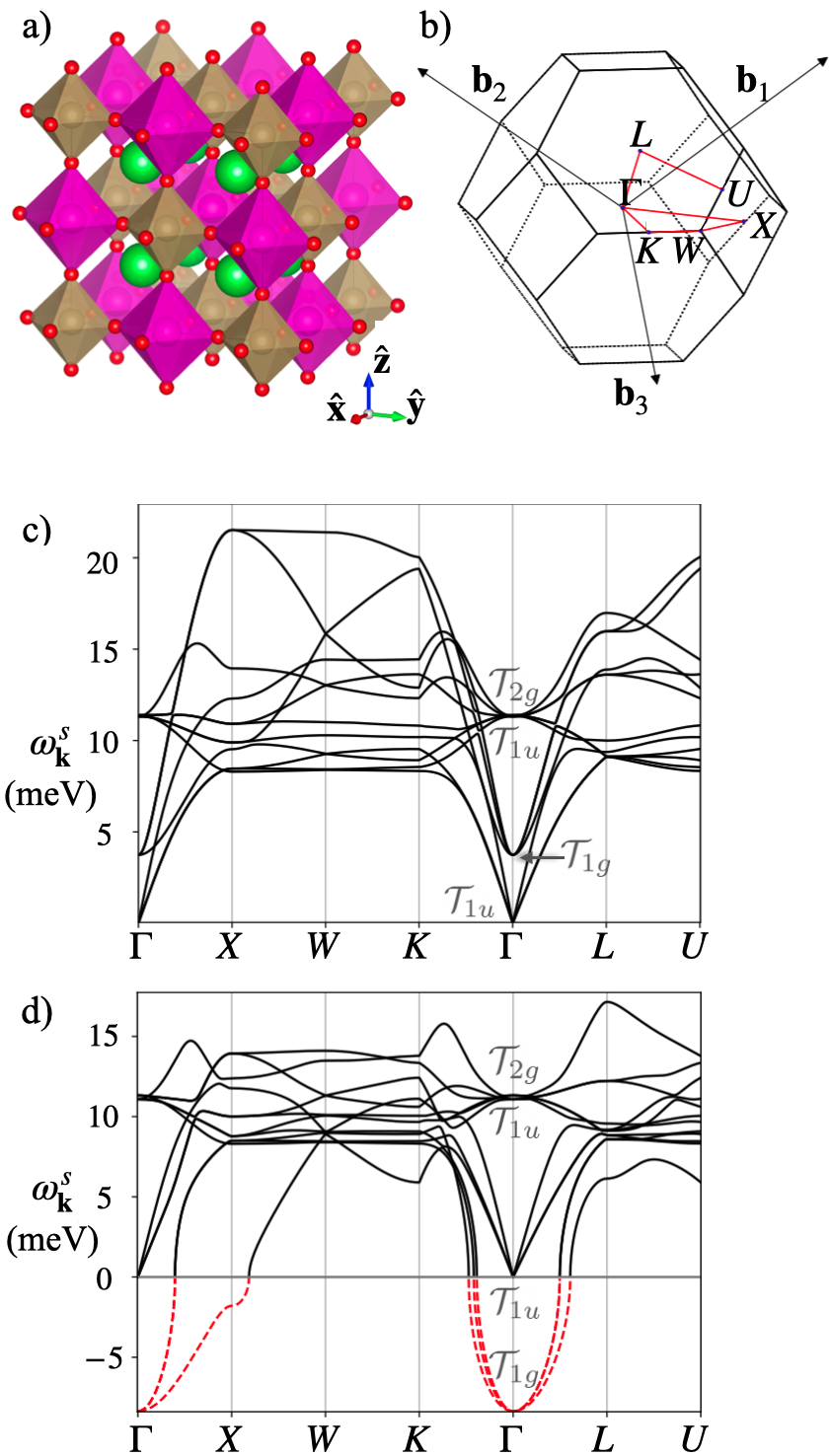}\caption{\label{fccPhonons} a) Ba$_{2}M$OsO$_{6}$ in the ideal rock-salt
structure with space group $Fm\bar{3}m$, where OsO$_6$ (light brown) and
$M$O$_6$ (fuschia) corner-sharing octahedra alternate in an fcc pattern surrounding Ba ions (green).
b) 1$^{\text{st}}$ Brillouin zone of the fcc primitive reciprocal
lattice vectors along with points of high symmetry. Low-energy phonon
spectrum for c) Ba$_2$CaOsO$_6$ and d) Ba$_2$CdOsO$_6$ in the $Fm\bar{3}m$ structure as calculated by the \emph{ab}-\emph{initio}
approach detailed in Appendix \ref{sec:appAabinit}; the phonon
irrep at the $\Gamma$ point is indicated in grey. The k-path traversed
is indicated in b). The presence of unstable $\mathcal{T}_{1g}$ modes for the
cubic $M=$ Cd material (dashed red) signals a structural distortion
instability induced by octahedral tilting at low temperatures. }
\end{figure}
Early approaches to analyze perovskite structural properties
include Goldschmidt's tolerance factor, which is adapted for
DPs $A_{2}BB'\text{O}{}_{6}$ as 
\begin{equation}
t=\frac{(r_{A}+r_{\text{O}})}{\sqrt{2}\left(\frac{r_{B}+r_{B'}}{2}+r_{\text{O}}\right)}\label{eq:gtf}
\end{equation}
where $r_{l}$ is the radius of ion $l$; $t=1$ would then correspond
to the high-symmetry cubic structure with space group symmetry $Fm\bar{3}m$ (no. 225) for the rock-salt formation, see Fig. \ref{fccPhonons}a)
\citep{Goldschmidt1926,babel1973}. Deviations from $t=1$ indicate a mismatch between the size of
the $A$ and $B/B'$ ions, inducing structural distortions that lower the crystal symmetry. The Ba$_{2}M$OsO$_{6}$ materials we
consider fall into two doppelg{\"a}nger pairs since the radii
of the Mg$^{2+}$/Zn$^{2+}$ (0.72/0.74 \AA) and Ca$^{2+}$/Cd$^{2+}$ (1.00/0.95 \AA) ions are roughly 
equivalent, resulting in tolerance factors of $t_{\text{Mg,\text{Zn}}}\sim1.04$
and $t_{\text{Ca,\text{Cd}}}\sim0.985$ \citep{rumble2017crc}. As
a result, one may expect the structural properties within each pair
of doppelg{\"a}ngers to be equivalent. 

We test this hypothesis by investigating
the phonon dispersion of each material in its electronic ground state,
which can be calculated using DFT. We
first perform a structural optimization where the ions are relaxed
from the initial $Fm\bar{3}m$ structure. The fcc conventional unit
cell has $\text{Os}$ at representative Wyckoff position $4a=\left(0,0,0\right)$,
$M$ at $4b=\left(\frac{1}{2},\frac{1}{2},\frac{1}{2}\right)$, $\text{Ba}$
at $8c=\left(\frac{1}{4},\frac{1}{4},\frac{1}{4}\right)$, and O at
$24e=\left(w,0,0\right),$ with $w\sim0.23-0.24$ depending on the
choice of $M$. We relax the primitive unit cell which has lattice
constant $a\sim5.7-6.0\text{ \AA}$ and contains one formula unit, ie. 10 atoms.
Once the optimized structure is obtained, we calculate the interatomic
force constants using density functional perturbation theory (DFPT)
\citep{Giannozzi1991,Gonze1997}, which, after Fourier interpolation,
yields the dynamical matrix at non-zero $\mathbf{q}$. Diagonalization
of the dynamical matrix then yields the dispersion $\omega_{s}\left(\mathbf{q}\right)$
for a given phonon branch $s=1,\ldots,30$. Computational details
of this three-step procedure are given in Appendix \ref{sec:appAabinit}. 
\begin{figure}
\centering{}\includegraphics[scale=0.26]{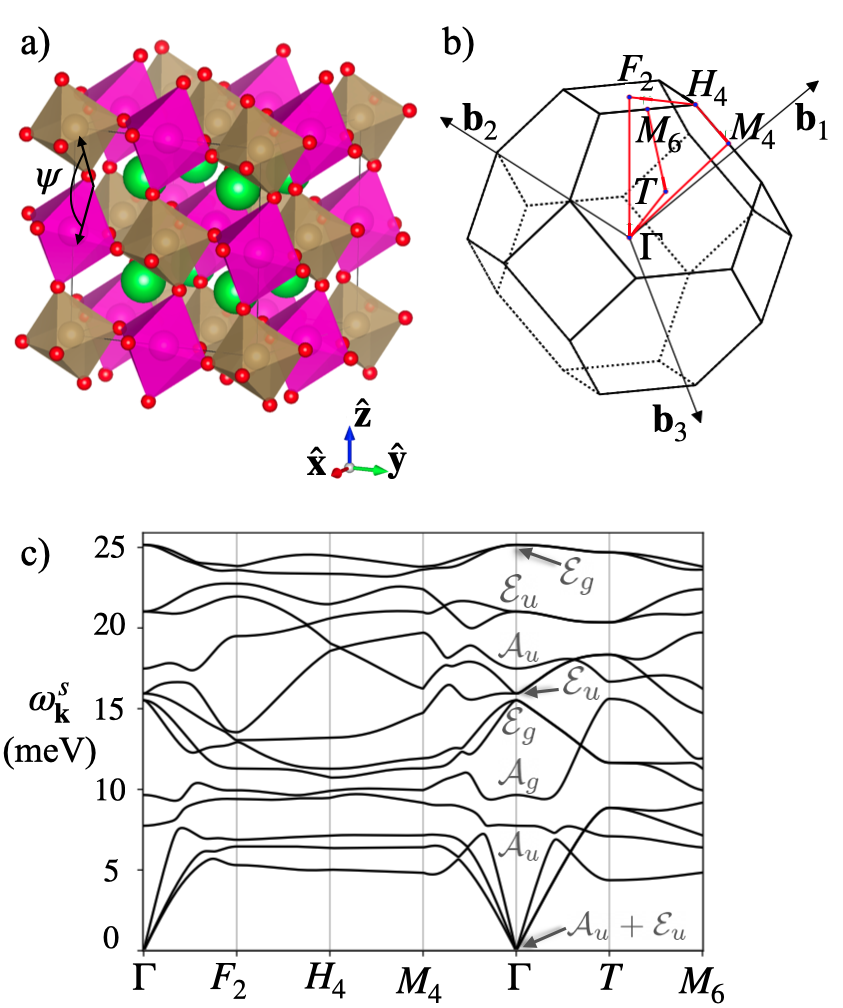}\caption{\label{trig} a) Crystal structure of the $a^{-}a^{-}a^{-}$ distorted
Ba$_{2}$CdOsO$_{6}$ with space group $R\bar{3}$, where each OsO$_6$
octahedron undergoes a uniform tilting in size and orientation. The
Cd-O-Os angles is given by $\psi=148.3^{\circ}$. b) 1$^{\text{st}}$
Brillouin zone of the rhombohedral primitive reciprocal lattice vectors
along with points of high symmetry; the rhombohedral setting is used.
c). Low-energy phonon spectrum for the crystal structure shown in
a) calculated by \emph{ab}-\emph{initio} techniques. The k-path traversed
is indicated in b). The lack of imaginary phonon frequencies indicates
that the $R\bar{3}$ structure is dynamically stable at low temperatures
for the $M=$ Cd compound.}
\end{figure}

The phonon dispersion for the Ca/Cd doppelg{\"a}nger pair in the $Fm\bar{3}m$
structure is shown in Fig. \ref{fccPhonons}; the Mg/Zn pair are given in Appendix \ref{sec:appAabinit} for completeness \footnote{Note added: In the process of producing this manuscript we became aware of Ref. \citep{voleti2022probing}'s study which also features the phonon dispersion for the $M=$ Ca material in the $Fm\bar{3}m$ structure. Our results match those of Ref. \citep{voleti2022probing} and includes the $M=$ Mg, Zn, Cd materials in addition.}. 
We will use calligraphic font to distinguish phonon irreps from the states in Eq. \eqref{eq:EgT2gcubic}, ie. the $E_g$ spin-orbital doublet vs. the $\mathcal{E}_g$ phonon mode.
At the $\Gamma$ point the 27 optical phonon modes decompose into irreps
of $O_{h}$ as $\mathcal{A}_{1g}\oplus \mathcal{E}_{g}\oplus \mathcal{T}_{1g}\oplus2\mathcal{T}_{2g}\oplus4\mathcal{T}_{1u}\oplus \mathcal{T}_{2u}$ \citep{Kroumova2003}. 
The clearest difference among the Ca/Cd pair is the $\mathcal{T}_{1g}$ mode which takes on an imaginary frequency at the $\Gamma$ point for the Cd compound, see Fig. \ref{fccPhonons}d). This occurs when the phonon spectrum is computed about an unstable structural equilibrium \citep{maradudin1971}; in this case, our \textit{ab}-\textit{initio} DFT analysis suggests that the ideal cubic structure of Ba$_2$CdOsO$_6$ with $Fm\bar{3}m$ symmetry is unstable and undergoes a structural transition to a state with lower symmetry at low temperatures.

The correct structure for Ba$_2$CdOsO$_6$ is obtained by displacing the atoms according to the three unstable eigenvectors associated with the $\mathcal{T}_{1g}$ modes. \textcolor{black}{This corresponds to fixing the positions of the Ba, Cd, and Os atoms while rotating the oxygen ions about the [111] direction.} The resulting structure is shown in Fig. \ref{trig}a) and consists of uniformly staggered octahedral canting, ie. $a^{-}a^{-}a^{-}$ in Glazer
notation with space group $R\bar{3}$ (no. 148) \citep{glazer1972classification, howard2003ordered}. The tilting reduces the Cd-O-Os
bond angle to $\psi = 148.3^{\circ}$ as shown in Fig. \ref{trig}a). While the space group symmetry is lowered, the primitive unit cell still contains one formula unit and maintains its shape; ie. rhombohedral with $\alpha=60^\circ$. We repeat the \emph{ab}-\emph{initio} analysis for Ba$_2$CdOsO$_6$ in the low-symmetry structure and find that all optical phonon modes carry non-zero frequency, see Fig. \ref{trig}c), indicating that the $R\bar{3}$ structure is stable. 
\section{$E_g$ and $T_{2g}$ States Under Trigonal Distortion\label{sec:trigDistort}}
In the previous section we argued that the stable structure for Ba$_2$CdOsO$_6$ has space group $R\bar{3}$, which breaks most $O_h$ point group operations at each Os$^6+$ ion yet retains the $C_3$ symmetry about the [111] axis. This is compatible with trigonal deformation of the OsO$_6$ octahedra; for Ba$_2$CdOsO$_6$ we find a compression of the OsO$_6$ cage along the [111] axis, see Fig. \ref{trigonalCEFocta} in Appendix \ref{sec:apptrigonalCEF}. In the trigonal environment the $t_{2g}$ orbital degeneracy is reducible into $t_{2g}=a_g \oplus e'_g$ irreps of $S_6$. The $E_g$ non-Kramers doublet arising from $J=2$ is not protected by the time-reversal symmetry and may be sensitive to the breaking of cubic crystalline symmetries. However, as the trigonal distortion preserves $C_3$ symmetry, the multipolar physics of the $E_g$ doublet may yet survive if the excited $T_{2g}$ states (which decompose as $T_{2g}=A_{g}\oplus E'_{g}$ in analogy with $t_{2g}$ orbitals) do not cross energies with the $E_g$ doublet.

\begin{figure}
\includegraphics[scale=0.19]{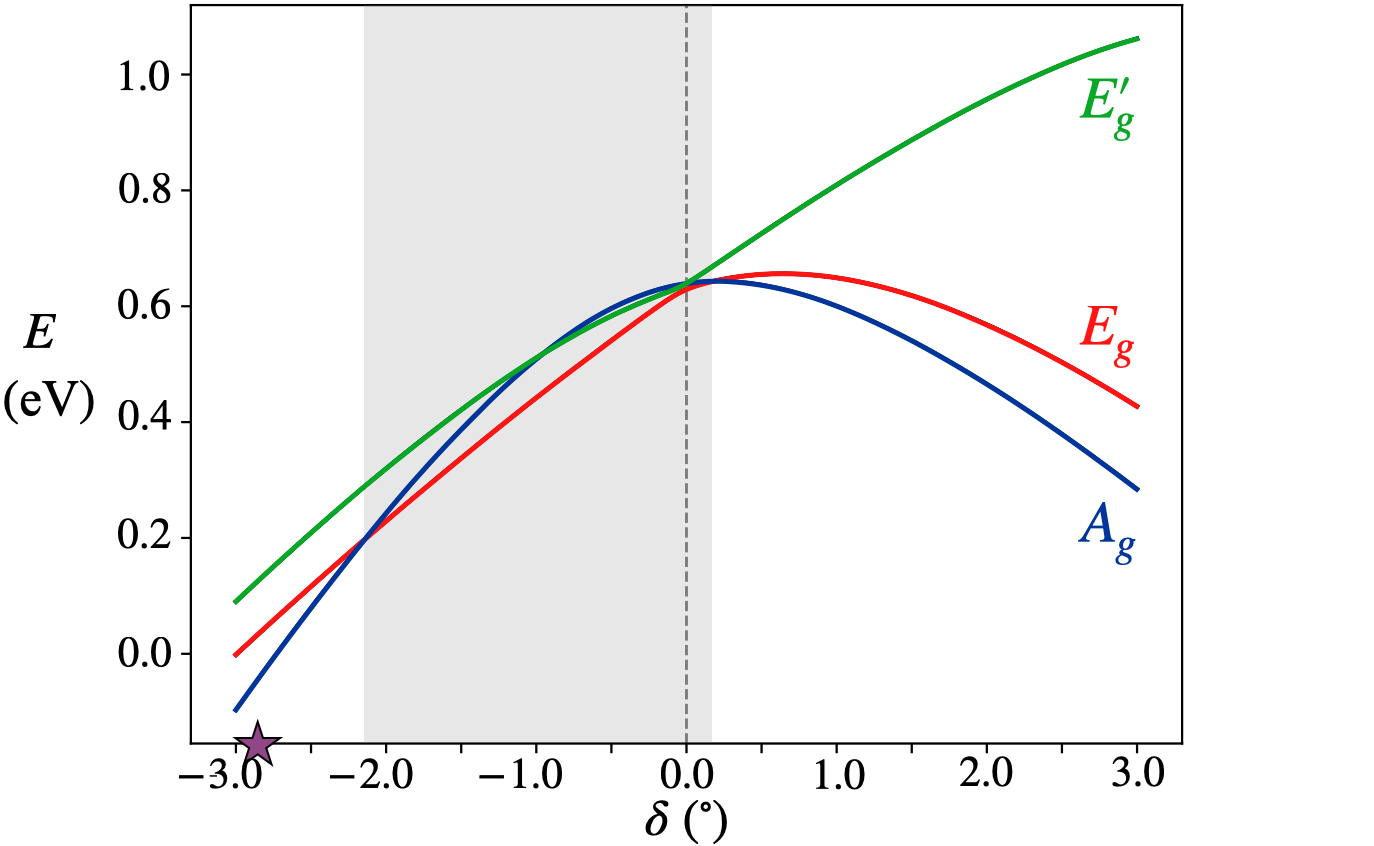}
\caption{\label{fig:EDtrigonal} ED calculation of the five lowest eigenvalues of $H_\text{int} + H_\text{SOC} + H_\text{CEF}(\delta)$ as $\delta$ is varied with $10Dq=4$ eV, $U=2.5$ eV, $\textcolor{black}{J_H}=0.5$ eV, and $\xi=0.4$ eV. $\delta>0$ ($\delta<0$) corresponds to trigonal elongation (compression). At $\delta=0$ the $T_{2g}$ triplet is separated by a gap of $\Delta \sim 10 $ meV from the $E_g$ doublet (red) and splits into an $A_g$ singlet (blue) and $E'_g$ doublet (green). The shaded region corresponds to the range of distortion where the $E_g$ doublet remains the lowest energy level. The purple star corresponds to the distortion in Ba$_2$CdOsO$_6$ of $\delta=-2.85^\circ$.}
\end{figure}

We investigate the fate of the multipolar physics by modelling the trigonal CEF and using exact diagonalization (ED) to solve for the single-ion spectrum of the electronic $5d^2$ configuration. The CEF Hamiltonian is given by 
\begin{equation}
H_\text{CEF}(\delta) = 10Dq \sum_{\alpha \sigma;\beta \sigma'} \left[\Xi^{\alpha \beta}(\delta) \otimes \mathbb{1}_{\sigma\sigma'}\right] d^\dagger_{\alpha \sigma} d_{\beta \sigma'},
\end{equation}
where $d_{\alpha \sigma}^{\dagger}$
creates a single electron with spin $\sigma\in\left\{ +,-\right\} $ in $d$ orbital $\alpha$, and $10Dq=4$ eV as is typical for the materials of interest \citep{Churchill_2022}. $\delta$ is an angle parameterizing the degree of trigonal distortion with $\delta > 0$ ($\delta<0$) corresponding to trigonal elongation (compression), see Fig. \ref{trigonalCEFocta}; the relaxed structure for Ba$_2$CdOsO$_6$ obtained in Sec. \ref{sec:StructuralOs} has $\delta=-2.87^\circ$, equivalent to a reduction of Os-O bonds by $\sim2\%$. The orbital level scheme is dictated by the $5\times 5$ matrix $\Xi$ which we estimate within the point-charge approximation in Appendix \ref{sec:apptrigonalCEF}. The local physics is also governed by SOC
\begin{equation}
H_\text{SOC} = \xi \sum_{\alpha \sigma;\beta \sigma'} \left[\mathbf{l}^{\alpha\beta}\cdot\mathbf{s}_{\sigma\sigma'} \right] d^\dagger_{\alpha \sigma} d_{\beta \sigma'},
\end{equation}
where $\mathbf{l}$ and $\mathbf{s}$ are angular momentum operators of $l=2$ and $s=1/2$ respectively and $\xi$ is the single-particle SOC strength, and the Kanamori-Hubbard interactions
\begin{align}
\label{eq:KHamiltonian}
H_{\text{int}} & =U\sum_{\alpha}n_{\alpha+}n_{\alpha-}\nonumber + U' \sum_{\alpha\neq\beta}n_{\alpha+}n_{\beta-}\nonumber \\
& +\left(U'-\textcolor{black}{J_H}\right)\sum_{\alpha<\beta,\sigma}n_{\alpha\sigma}n_{\beta\sigma}\\
 & -\textcolor{black}{J_H}\sum_{\alpha\neq \beta}\left(d_{\alpha+}^{\dagger}d_{\alpha-}d^{\dagger}_{\beta-}d_{\beta+}+d_{\alpha+}^{\dagger}d_{\beta-}d_{\alpha-}^{\dagger}d_{\beta+}\right), \nonumber
\end{align}
where $n_{\alpha \sigma}=d_{\alpha \sigma}^{\dagger}d_{\alpha \sigma}$, \textcolor{black}{$J_H$} is Hund's coupling strength, and $U$ $\left(U'=U-2\textcolor{black}{J_H}\right)$ is the intraorbital (interorbital) Hubbard parameter. For a given $\delta$ the local Hamiltonian $H_\text{int} + H_\text{SOC} + H_\text{CEF}(\delta)$ can be diagonalized within the basis of $\binom{10}{2}=45$ possible $d^2$ states with $10Dq=4$ eV, $U=2.5$ eV, $\xi=0.4$ eV, and finite \textcolor{black}{$J_H$}. At $\delta=0$ the five lowest states are the $E_g$ doublet and $T_{2g}$ triplet and we choose $\textcolor{black}{J_H}=0.2U$ so that the $E_g$-$T_{2g}$ splitting is $\Delta \sim 10$ meV. \textcolor{black}{For our choice of $U$ this corresponds to $J_H=$ 0.5 eV which is a reasonable upper bound for the Hund's coupling in the $5d$ series \cite{khomskii2014}}. The result for finite $\delta$ is given in Fig. \ref{fig:EDtrigonal}; within the set of parameters considered the five lowest eigenvalues are gapped from the rest of the spectrum.  

Unlike the tetragonal case, trigonal distortions do not split the non-Kramers $E_g$ doublet due to the presence of $C_3$ symmetry, while the excited triplet decomposes as $ T_{2g} = A_g\oplus E'_g$. The $E_g$-$E'_g$ doublets undergo level repulsion with the $E_g$ states lower in energy for all $\delta$. On the other hand,  the $A_g$ singlet is the lowest eigenvalue at larger distortions, crossing the $E_g$ doublet and suppressing the multipolar physics within. In the case of elongation, $A_g$ is lower in energy than $E'_g$ and the singlet crosses the $E_g$ doublet at a relatively small distortion of $\delta\sim+0.18^\circ$. The case of trigonal compression is more interesting, as the $E_g$ doublet remains the state of lowest energy until $\delta\sim-2.1^\circ$. 

The asymmetry between compression and elongation can be traced back to filling the $a_g$ and $e'_g$ single-electron orbital states. Trigonal elongation lowers the $e'_g$ doublet and the resulting two-electron state has its orbital degrees of freedom quenched. On the other hand, trigonal compression lowers the $a_g$ singlet, and the two-electron state is predominantly composed of a) the filled singlet and b) the state where both $a_g$ and $e'_g$ are singly-occupied. Thus there is a competition between the distortion and Hund's coupling, with the latter seeking to eliminate orbital fluctuations. This frustration manifests itself in the relatively small energy difference between the $E'_g$ and $A_g$ states when $\delta\lesssim0$ (see Fig. \ref{fig:EDtrigonal}), and is finally resolved in the limit of strong distortions by stabilizing the $A_g$ singlet. The trigonal compression in Ba$_2$CdOsO$_6$ is given by $\delta=-2.85^\circ$ which is just beyond the $A_g$-$E_g$ crossing. This suggests that the multipolar physics in rhombohedral Ba$_2$CdOsO$_6$ is likely to be revealed in pressure-tuned experiments by approaching the cubic limit.

\section{Summary and Discussion}
In summary, we investigated the structural properties of four $5d^2$ Os DPs Ba$_2M$OsO$_6$ using an \textit{ab}-\textit{initio} calculation of the phonon spectrum. We found that while the $M=$ Mg, Zn, Ca compounds have a stable cubic structure with space group $Fm\bar{3}m$, the $M=$ Cd material forms the rhombohedral $R\bar{3}$ structure at low temperatures. The transition to the low-symmetry state is  indicated by softening of the $\mathcal{T}_{1g}$ phonon modes which results in out-of-phase tilting of the oxygen octahedra in each of three crystallographic directions, ie. $a^-a^-a^-$. The rhombohedral structure generally allows for trigonal distortion of the OsO$_6$ octahedra, which modifies the physics of the $J=2$ states by splitting the excited $T_{2g}$ triplet into an $E'_g$ doublet and $A_g$ singlet. There exists a finite window where the $E_g$ doublet is the lowest energy level, allowing for the ordering of various multipolar orders. A strong enough deformation stabilizes the $A_g$ singlet and suppresses the $E_g$ doublet's multipolar magnetism.

It is interesting to note that while $r_\text{Ca} > r_\text{Cd}$ by only 0.05 \AA, Ba$_2$CaOsO$_6$ remains cubic at low temperatures despite having a lower tolerance factor than Ba$_2$CdOsO$_6$. This is likely due to the non-ionic nature of DPs which cannot be accounted for by the tolerance factor calculated from tabulated ionic radii \citep{vasala2015}. One likely explanation for the difference between the Ca and Cd compounds is the number of $d$-electrons present on the $M^{2+}$ ion, ie. $d^0$ and $d^{10}$ respectively. The capacity for $\pi$-bonding in Ba$_2$CdOsO$_6$ favours the reduction of the Cd-O-Os bond angles, while the lack of such for Ba$_2$CaOsO$_6$ results in linear Ca-O-Os bonds \citep{SAINES2007401,SAINES20072991,SAINES20073001}. While Eq. \eqref{eq:gtf} provides a good first approach to the classification of perovskite structural properties, the fact that Ba$_2$CaOsO$_6$ and Ba$_2$CdOsO$_6$ form a doppelg{\"a}nger pair but have different low-temperature structures motivates the need to go beyond the tolerance factor by considering each compound's chemical properties.

\textcolor{black}{The presence of trigonal distortion allows for new interorbital exchange pathways including $xy$-$xz/yz$ hopping (parameterized by $t_4$ in Ref. \cite{Churchill_2022}) for bonds parallel to the $xy$-plane. The parameters of the $E_g$ pseudospin-1/2 effective model are modified in the weak hopping limit, and the dependence on $t_4$ in $5d^2$ DPs with small trigonal distortion is an interesting subject left for future study. We find that the distortion in Ba$_2$CdOsO$_6$ is large enough to place the material outside the region where the effective $E_g$ model is valid.} Since the singlet crosses the doublet and becomes the lowest state separated by a small gap from the $E_g$, the intersite exchange interaction may become important in Ba$_2$CdOsO$_6$. 
While the singlet itself has no moment, the intersite exchange interaction may induce a magnetic moment.

Our first main conclusion is that Ba$_2$CdOsO$_6$ should feature a cubic-rhombohedral structural transition at low temperatures. Investigation of this scenario can be determined using high-resolution synchrotron x-ray diffraction, a reliable technique for detecting antiferro-distortions in $5d$ DPs \citep{hirai2020,Maharaj_2020}. The cubic-rhombohedral structural transition has not been considered in previous studies of the $5d^2$ Ba$_2M$OsO$_6$ series yet is common in other DPs, for example the strontium osmate Sr$_{2}$CrOsO$_{6}$ \citep{KROCKENBERGER20071854, krockenberger2007, MORROW201646}
and a series of strontium antimony oxides Sr$_{2}B''$SbO$_{6}$ \citep{FAIK20081759, FAIK20092656, ORAYECH2017265}. 
The structural properties of these materials has been explored using x-ray and neutron powder diffraction measurements, and display a series of structural transitions
(monoclinic $\rightarrow$ rhombohedral $\rightarrow$ cubic) as temperature
is increased. In particular, the $R\bar{3}\rightarrow Fm\bar{3}m$
transition is observed to be of second-order as predicted by Landau theory \citep{howard2003ordered} and corresponds to the suppression of the cubic $(642)$ peak below the ordering temperature \citep{FAIK20081759, FAIK20092656, ORAYECH2017265, orayech2015}. Similar observations in the low-temperature phase of Ba$_2$CdOsO$_6$ would serve as experimental evidence for rhombohedral symmetry in this material. Another signature of the symmetry lowering would be the splitting of triplet phonon modes $\mathcal{T}_{1g,2g}$ at the $\Gamma$ point into irreps of $S_6$ as $\mathcal{T}_{1g,2g} = \mathcal{A}_g \oplus \mathcal{E}_g$. In Fig. \ref{trig}c) we see that the $\mathcal{A}_g$-$\mathcal{E}_g$ splitting is roughly 5 meV; such an energy difference can be resolved via Raman scattering. \textcolor{black}{Moreover, inelastic scattering experiments can reveal signatures of strong vibronic coupling in the excitation spectrum as recently demonstrated for K$_2$IrCl$_6$ \citep{iwahara2023}. In this case, the spin-orbit-lattice entangled vibronic state does not result in static structural distortions due to dynamic Jahn-Teller considerations; we leave this scenario open for future research and analysis.}
Our findings highlight the importance of considering the interplay between crystal structure and electronic states when exploring the magnetic properties of $5d$ spin-orbit coupled magnets. 

The second main conclusion is that multipolar physics can survive in $5d^2$ materials with trigonal distortions. The $E_g$ doublet does not split under trigonal distortions and there exists a window of finite distortion where it remains the lowest energy level, especially for the case of trigonal compressions. This suggests that multipolar physics can be found in structures with lower symmetry such as $R\bar{3}$. Interestingly, this space group is common amongst several honeycomb compounds including CrI$_3$ \citep{morosin1964}. A recent proposal has suggested that $5d^2$ honeycomb compounds could host exotic ordered and disordered multipolar phases, including the Kitaev multipolar liquid \citep{Rayyan2023}. The identification of a $5d^2$ honeycomb material with, e.g., $R\bar{3}$ symmetry and small trigonal compression is an excellent first step to the realization of the Kitaev multipolar liquid and forms an interesting direction for future work.

\begin{acknowledgments}
A.R. is grateful for helpful discussions with D. Churchill and S. Voleti. 
We acknowledge support from the Natural Sciences and Engineering Research Council of Canada Discovery Grant No. 2022-04601. 
H.Y.K. also acknowledges support from the Canadian Institute for Advanced Research and the Canada Research Chairs Program. 
Computations were performed on the Niagara supercomputer at the SciNet HPC Consortium. SciNet is funded by: the Canada Foundation for Innovation; the Government of Ontario; Ontario Research Fund - Research Excellence; and the University of Toronto.
\end{acknowledgments}

\appendix

\section{\emph{ab}-\emph{initio} Calculations and Phonon Dispersions for $M=\text{Mg/Zn}$\label{sec:appAabinit}}

\begin{figure}
\includegraphics[scale=0.3]{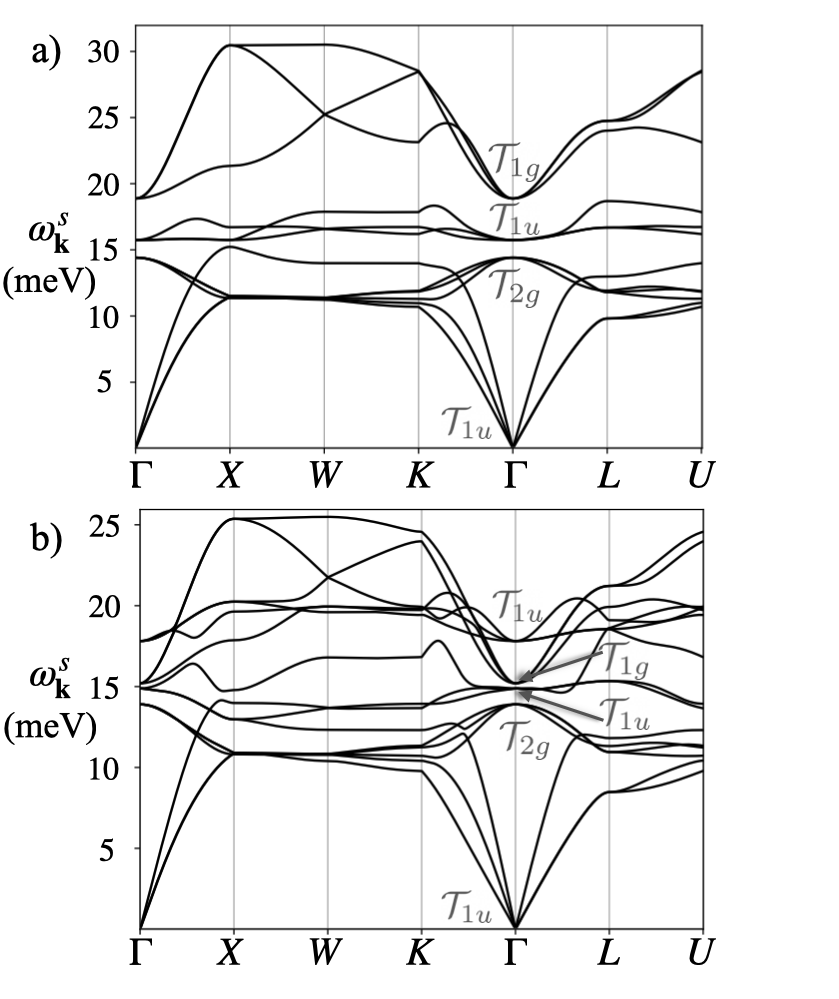}
\caption{\label{phononsMgZncubic} Low-energy phonon spectrum calculated via
an \emph{ab}-\emph{initio} approach for the a) Ba$_2$MgOsO$_6$ and
b) Ba$_2$ZnOsO$_6$. materials in the $Fm\bar{3}m$ structure, with symmetry
points shown in Fig. \ref{fccPhonons}. The phonon irrep at the $\Gamma$
point is shown in grey. The lack of imaginary phonon frequencies
indicates the stability of the $Fm\bar{3}m$ structure at low temperatures for the $M=\text{Mg}, \text{Zn}$ compounds.}
\end{figure}

The DFT calculations are performed using Vienna Ab initio Simulation
Package (VASP) with the Perdew-Burke-Ernzerhof exchange-correlation
functional and cutoff energy of 600 eV \citep{kresse1996,perdew1996}.
The crystal structures of all four materials considered in this work
are obtained from the Materials Project \citep{Jain2013}. Structural
relaxation of the primitive unit cell in the initial $Fm\bar{3}m$
structure is performed using a $\Gamma$-centered $8\times8\times8$
k-mesh with maximum force/atom of $10^{-4}$ eV/\AA. In the
ionic relaxation all degrees of freedom are varied including ionic
positions, cell shape, and cell volume; care is taken to eliminate
artifacts associated with the Pulay stress by repeated re-relaxation
of the intermediate ionic positions. We then used VASP to calculate
interatomic force constants using DFPT simulations on a $2\times2\times2$
supercell with a $\Gamma$-centered $4\times4\times4$ k-mesh and
an energy convergence threshold of $10^{-8}$ eV. The supercell contains
8 formula units and has a (super)lattice constant of roughly $12$ \AA;
a large supercell is chosen to eliminate artifacts arising from long-range
forces. Finally, phonopy performs the diagonalization of the dynamical
matrix at $\mathbf{q}$ commensurate with the supercell and the Fourier
interpolation at general $\mathbf{q}$ \citep{TOGO20151,Togo2023}. In Fig. \ref{phononsMgZncubic} we present the phonon dispersion for the Mg/Zn doppelg{\"a}nger pair.
VASPKIT was used to prepare the VASP simulations such as in creation
of the position and ionic relaxation/DFPT input files \citep{WANG2021108033}.
Crystal structures shown in Figs. \ref{fccPhonons}a) and \ref{trig}a)
were visualized using VESTA \citep{momma2008}. The Brillouin zones
in Fig. \ref{fccPhonons}b) and \ref{trig}b) were generated using
the Atomic Simulation Environment package \citep{Larsen_2017} with
symmetry points labelled according to the HPKOT convention \citep{HINUMA2017140}. 

\section{Construction of the trigonal CEF\label{sec:apptrigonalCEF}}
\begin{figure}
\centering{}\includegraphics[scale=0.20]{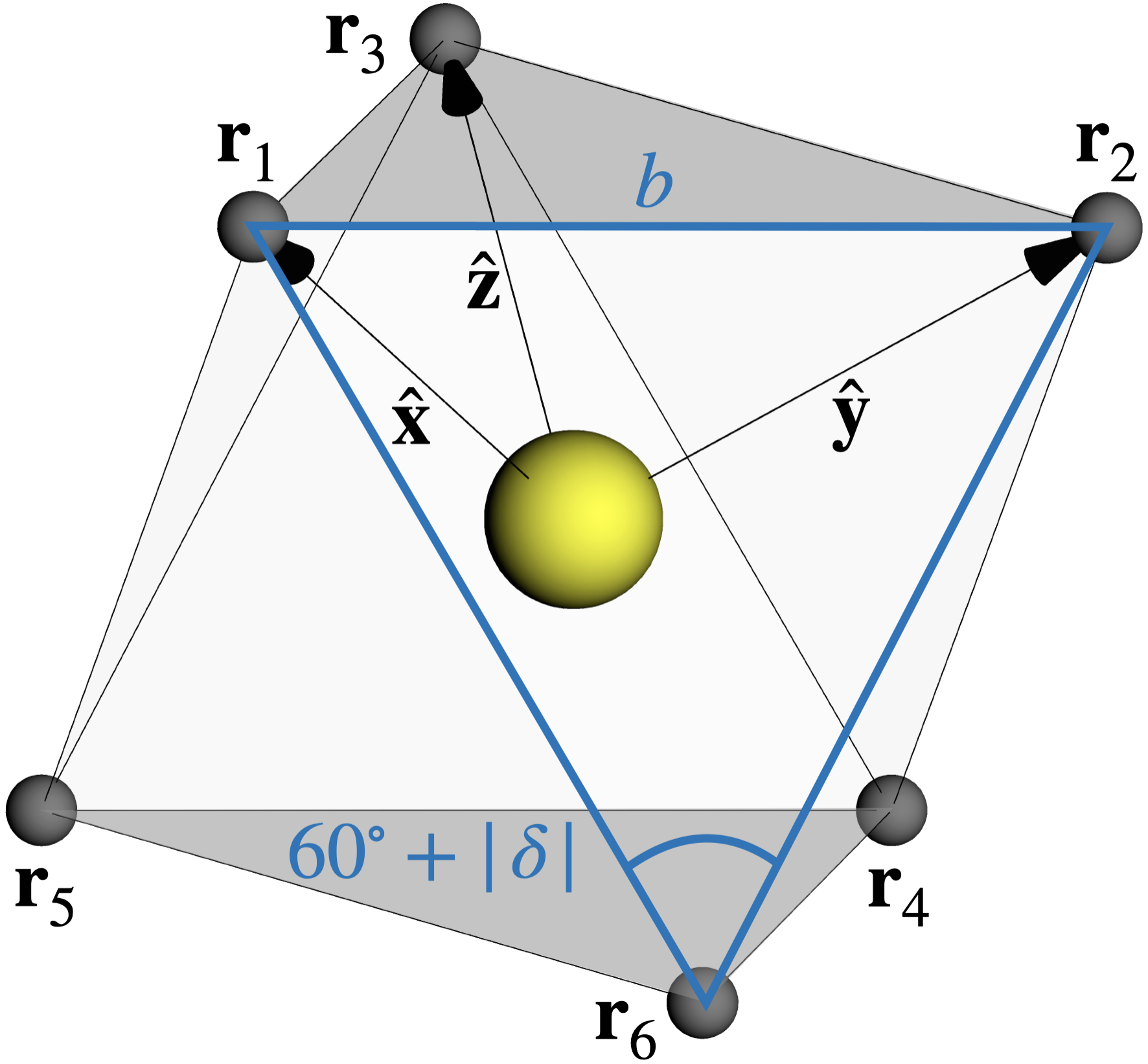}
\caption{\label{trigonalCEFocta} Geometry used to derive the CEF in the case of trigonal compression $(\delta < 0)$, where the $[111]$ and $[\bar{1}\bar{1}\bar{1}]$ faces (shaded) are pressed towards each other. In Ba$_2$CdOsO$_6$ the three oxygen ions at $\mathbf{r}_{1,2,6}$ form an isosceles triangle (blue) with base $b=2.94$ \AA $\,$and $\delta=-2.85^\circ$. The $xyz$ coordinates in this figure (ie. the local octahedral coordinates at each Os atom) differ from the $xyz$ coordinates shown in Fig. \ref{trig}a) (the crystallographic directions) by a rotation about the [111] direction due to the octahedral canting in the $R\bar{3}$ structure.}
\end{figure}

We can estimate the effect of trigonal distortion on electrons in the $5d$ shell by constructing a trigonal CEF Hamiltonian in the point charge limit. We will only consider trigonal compression and elongation of the OsO$_6$ cage, but other deformations are allowed in principle \citep{stavro2021}. The oxygen positions can be parameterized as 
\begin{align}
\label{eq:oxygenpositions}
\mathbf{r}_{1} & =\frac{b}{\sqrt{2}}\hat{\mathbf{x}}+f(\delta)\hat{\mathbf{c}}, & \mathbf{r}_{4} & =-\mathbf{r}_{1},\nonumber \\
\mathbf{r}_{2} & =\frac{b}{\sqrt{2}}\hat{\mathbf{y}}+f(\delta)\hat{\mathbf{c}}, & \mathbf{r}_{5} & =-\mathbf{r}_{2},\\
\mathbf{r}_{3} & =\frac{b}{\sqrt{2}}\hat{\mathbf{z}}+f(\delta)\hat{\mathbf{c}}, & \mathbf{r}_{6} & =-\mathbf{r}_{3},\nonumber 
\end{align}
where $\hat{\mathbf{c}} = \frac{1}{\sqrt{3}}\left(\hat{\mathbf{x}}+\hat{\mathbf{y}}+\hat{\mathbf{z}}\right)$, $b$ is the length of the octahedron edge in the ideal limit, and  
\begin{equation}
f(\delta) = \frac{b}{4}\left[\text{tan}\left(60^\circ+\frac{\delta}{2}\right) - \text{tan}\left(60^\circ\right)\right]
\end{equation}
where $\delta > 0\;(\delta < 0)$ corresponds to the case of trigonal elongation (compression), see Fig. \ref{trigonalCEFocta}. 
For Ba$_2$CdOsO$_6$ we find that $b=2.94\text{ \AA}$ and $\delta=-2.85^\circ$, equivalent to a reduction in Os-O bonds by $\sim2$\%.
Note that the $xyz$ coordinates in Eq. \eqref{eq:oxygenpositions} refer to the local octahedral coordinates at each osmium atom, which differ from the crystallographic $xyz$ coordinates by a rotation about the [111] direction. This is due to the octahedral tilting within the $R\bar{3}$ structure which affects all OsO$_6$ octahedra uniformly. With this in mind, we will continue to use the $xyz$ symbols to simplify notation.

The potential energy of the $5d^2$ electrons in each Os$^{6+}$ ion in the presence of the O$^{2-}$ point charges can be written in a multipole expansion given by
\begin{align}
\label{eq:multipolePotential}
V\left(\mathbf{r}\right) =\sum_{k=0}^{\infty}\sum_{p=-k}^{k}\;r^{k}\left[\frac{4\pi A}{2k+1}\sum_{i=1}^{6}\frac{Y_{kp}\left(\theta_{i},\phi_{i}\right)^{*}}{r_{i}^{k+1}}\right]Y_{kp}\left(\theta,\phi\right),
\end{align}
for $r\equiv|\mathbf{r}|<|\mathbf{r}_{i}|$ \citep{arfken2012mathematical}. The proportionality constant $A>0$ is fixed so that the $t_{2g}$-$e_g$ splitting at $\delta=0$ is $10Dq=4$ eV. We then evaluate matrix elements of Eq. \eqref{eq:multipolePotential} between hydrogenic wave functions $R_{n=5,l=2}(r)Y_{l=2,l_z}(\theta,\phi)$. 
Since $l=2$ we may restrict the $k$ summation to $k\leq4$ and we further ignore the $k=0$ contribution as it does not carry an angular dependence. When $\delta\neq0$ the $t_{2g}$ degeneracy is lifted and it becomes more appropriate to write the trigonal CEF in the following basis of single-electron creation operators \citep{sugano2012multiplets}
\begin{align}
d^\dagger_{a_{g}}&=\frac{1}{\sqrt{3}}\left(d^\dagger_{xy}+d^\dagger_{yz}+d^\dagger_{xz}\right),\nonumber \\
d^\dagger_{e_g'^{-}}&=\frac{1}{\sqrt{3}}\left(d^\dagger_{xy}+\omega^2 d^\dagger_{yz}+ \omega d^\dagger_{xz}\right),\nonumber \\
d^\dagger_{e_g'^{+}}&=\frac{-1}{\sqrt{3}}\left(d^\dagger_{xy}+\omega d^\dagger_{yz}+\omega^2 d^\dagger_{xz}\right),\label{eq:a1geg'basis} \\
d^\dagger_{e_g^{\mp}}&=\frac{\pm1}{\sqrt{2}}\left(d^\dagger_{z^2}\mp id^\dagger_{x^2-y^2}\right),\nonumber
\end{align}
where $\omega=\text{exp}\left(2\pi i/3\right)$ and the spin index is suppressed. For $\delta=-2.85^\circ$ as in Ba$_2$CdOsO$_6$ we find that the trigonal CEF is given by
\begin{equation}
\Xi^{\alpha\beta}(\delta)=\left(\begin{array}{c|cc|cc}
-0.197 & 0 & 0 & 0 & 0\\
\hline0 & 0.0406 & 0 & 0.145 & 0\\
 0 & 0 & 0.0406 & 0 & 0.145\\
\hline0 & 0.145 & 0 & 1.058 & 0\\
0 & 0 & 0.145 & 0 & 1.058
\end{array}\right),\label{eq:crystalFieldHamiltonian}
\end{equation}
and the numerical values are in units of $10Dq$. By diagonalizing Eq. \eqref{eq:crystalFieldHamiltonian} one can check that the trigonal compression $\delta<0$ yields a low-lying $a_{g}$ singlet whereas elongation $\delta>0$ prefers the lowering of the $e'_g$ doublet. 
Eq. \eqref{eq:crystalFieldHamiltonian} demonstrates the mixing of $e_g$ and $e_g'$ states of the same parity under trigonal deformations. 


%

\end{document}